\documentstyle{mn}
\input{epsf}
\begin{document}
\def\p0437{PSR J0437$-$4715}

\title[No Evidence for Coherent Diffractive Radiation Patterns in \p0437]
{No Evidence for Coherent Diffractive Radiation Patterns in the Millisecond Pulsar \p0437}

\author[M. Vivekanand]{M. Vivekanand \\	
National Center for Radio Astrophysics, TIFR, \\
Pune University Campus, P. O. Box 3, 
Ganeshkhind, Pune 411007, India. (vivek@ncra.tifr.res.in)}

\date{}

\maketitle
\begin{abstract}
An independent analysis of the 326.5 MHz data obtained by Ooty Radio Telescope
reveals no evidence for coherent diffractive radiation patterns in the 
millisecond pulsar \p0437.
\end{abstract}
\begin{keywords}
pulsars -- \p0437 -- single pulses -- radio -- radiation mechanism.
\end{keywords}
\section{Introduction}

Ables et al \cite{AMDV} (henceforth AMDV) claimed that they observed periodic
variation in the distribution of the positions (or phases) of the spiky emission 
in the milli-second pulsar \p0437. Their data were obtained at 326.5 MHz with the
Ooty Radio Telescope (ORT) using the incoherent de-dispersion technique; only a
single linear polarization was available. The sampling interval of their data was 
102.4 micro seconds ($\mu$s), but they apparently discovered a much smaller 
period of $\approx$ 20 $\mu$s for the fringes in the distribution of positions 
of spikes.  They state they could do so because ``in these distributions there 
is almost no noise in the traditional sense'', so they assumed the typical rms 
error on the positions of the spikes to be about 1.7 $\mu$s (``Gaussian of 4 
$\mu$s FWHM''); this is much smaller than either the sampling interval or the 
fringe period. They state that the above was evidence for ``diffraction fringes 
associated with coherent emission from a finite aperture'' on the pulsar.  

This was refuted by Jenet et al \cite{JAKPU}, who analyzed \p0437 data obtained 
at 1380 MHz with the Parkes Observatory. Using the coherent de-dispersion 
technique, they obtained a time resolution of 0.32 $\mu$s, which is much shorter 
than even the expected $\approx$ 4.7 $\mu$s fringe spacing at 1380 MHz, if the 
fringes are due to diffractive coherent radiation as claimed by AMDV.

This paper reports the result of an independent analysis on the data of AMDV.
No evidence is found for the fringes reported by them. The possible reasons for 
their result are discussed.

\section{Observations and data reduction}

The instrument, the method of observation, and the several variations of data
reduction for diverse purposes are given in AMDV, Vivekanand et al \cite{VAM98}
and Vivekanand \cite{MV20}. 

\begin{figure}
\epsfxsize=8.5cm \epsfbox{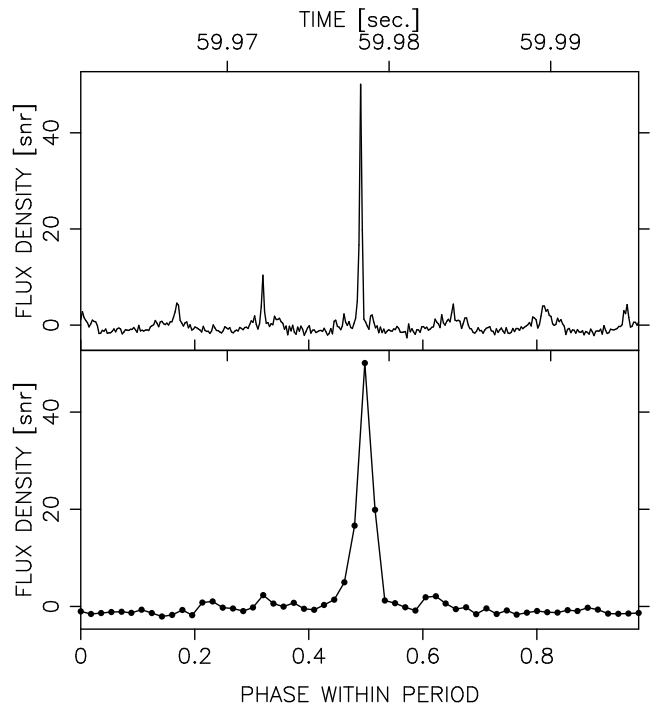}
\caption{
	 Similar to Fig.~1 of AMDV. {\bf Top panel}: Six periods of data 
	 centered on the strongest spike observed in \p0437 at 326.5 MHz using 
	 ORT (file labeled 50681546). The abscissa is in seconds measured from 
	 the start of the data file, while the ordinate is in units of the 
	 signal to noise ratio. {\bf Bottom panel}: The period containing the 
	 spike; the abscissa is in units of the period (this is also known as 
	 phase or longitude within the period).
	}
\label{fig1}
\end{figure}

The top panel of fig.~\ref{fig1} shows six periods centered around the highest 
spike in the data file labelled 50681546, observed at UT 15:46:18 on 9 Mar 1995, 
with a sampling interval of 102.4 $\mu$s. This file contains one of the highest 
sensitivity data available.  The ordinate was obtained in units of the signal to 
noise ratio (snr) in the following manner: 

First, the 681\,984 continuous time samples of pulsar flux data in the file 
were loaded into an array of length 1\,048\,576, and an FFT was performed to 
obtain the power spectrum. The most prominent features were the pulsar 
fundamental at 173.688427 Hz and 27 of its harmonics. Next were the second, 
fourth and sixth harmonics of the power line frequency of $\approx$ 48 Hz (this 
number varies by $\pm 1$ Hz at ORT depending upon the day and time of 
observation). At each of these 28 + 3 = 31 spectral features, a width of 1.0 Hz 
was filtered out. The rest, barring a few small spikes, looked like the 
spectrum of random noise, and was assumed to be due to receiver noise, 
intrinsic pulse to pulse flux variations of \p0437, etc. The mean value of this 
power spectrum is related to the root mean square (rms) random noise $\sigma$ 
in the time domain, by the generalized Perseval-Rayleigh theorem for digitized 
signals (Bracewell \cite{RB}). This $\sigma$ value was used to normalize the 
original time series, so that it is now in units of the snr.

\begin{figure}
\epsfxsize=8.5cm \epsfbox{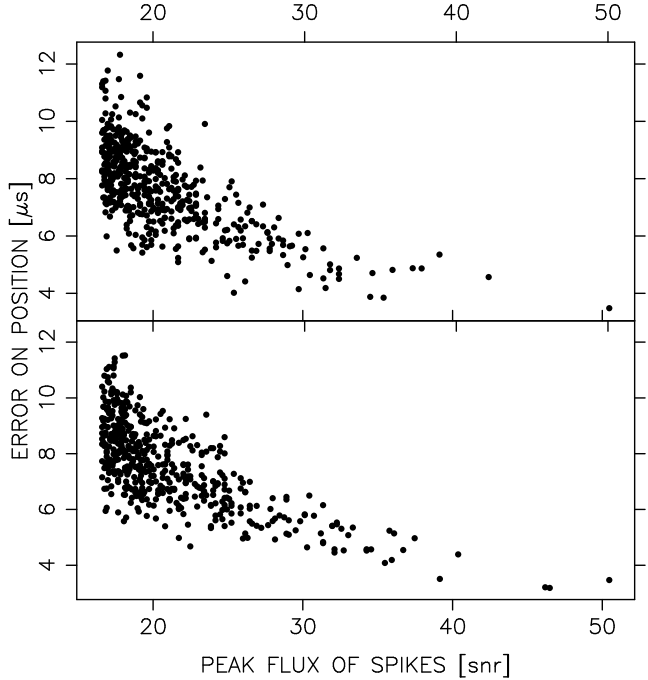}
\caption{
	 The height of 500 spikes (in units of signal to noise ratio) plotted
	 against the estimated root mean square error on their positions (in 
	 micro seconds). The top panel is for data file 50681546 while the
	 bottom panel is for data file 50681549. The expected anti-correlation 
	 is clearly visible.
	 }
\label{fig2}
\end{figure}

How does this snr compare with that obtained by AMDV? A direct comparison is
not possible since AMDV mention only the flux density of their spikes and not
their snr; they also do not specify the two data files they worked with. 
However in the caption of their fig.~1 they state that ``These spikes are 
about a third of the height of the largest seen in the whole data set''; 
since the largest spike in their fig.~1 has a height $\sol$ 100 Jansky (Jy), 
one can assume that the height of their largest spike was $\sol$ 300 Jy. Now, 
the theoretically expected $\sigma$ for the total power mode of observation 
at ORT for an 8 MHz (effective) bandwidth and a sampling interval of 102.4 
$\mu$s is 2.8 Jy, at the best phased condition of ORT; the corresponding 
number for the phase switched mode of observation is $\sqrt{2}$ times larger
(see Vivekanand \cite{MV95}). So the theoretical best snr to be expected by
AMDV for their highest spike is $\approx$ 300/2.8 $\approx$ 107, which is
about twice the number in fig.~\ref{fig1} here. This difference should reduce
if one takes into account the noise due to the pulse to pulse flux 
variations of \p0437, reduction in snr due to instrumental effects such
as digitization, de-phasing of ORT, etc.

This author feels that it is the $\sigma$ of this paper that should be
operative in estimating the snr of the spikes. However, the later analysis 
was also repeated with an effective $\sigma$ that is four times smaller, 
i.e., with an average snr that is at least two times better than the 
presumed snr of AMDV, with no change in the essential results.

Next, the positions of the highest 500 spikes in the data file, along with their
formal errors, were obtained as discussed in Appendix A. Figure~\ref{fig2} shows
the plot of these errors versus the height of the peaks, for data files labelled
50681546 and 50681549; the latter was observed at UT 15:49:39 on 9 Mar 1995 
(just 3.35 minutes after the former data file) with the same sampling interval 
(102.4 $\mu$s), and represents yet another file containing very high sensitivity 
data. The positional errors are inversely correlated with heights, as expected 
from Appendix A. While the highest peak in both files has a positional error of 
$\approx$ 3.5 $\mu$s, the mean positional error in the two files is $\approx$ 7.5 
$\mu$s with standard deviation of $\approx$ 1.5 $\mu$s. This is due to the rapidly 
decreasing number of spikes with increasing spike height. 

The positions obtained here are consistent with those of Vivekanand \cite{MV20}. 
The main differences are (1) the former are obtained from the original time series 
data, while the latter were obtained from the re sampled and folded data, and (b) 
the former are obtained by a weighted mean, while the latter were obtained by 
fitting Gaussian to the spikes.

\begin{figure}
\epsfxsize=8.5cm \epsfbox{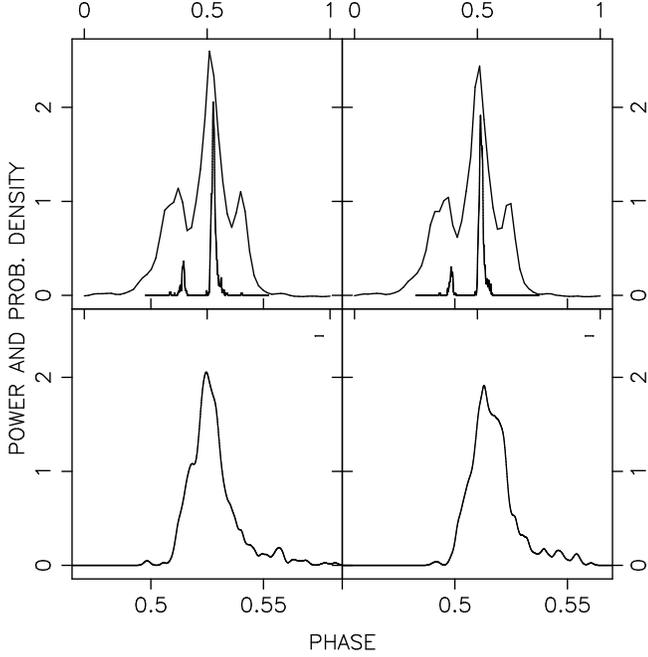}
\caption{
	 Similar to Fig.~2 of AMDV. {\bf Top panels}: The integrated profiles of data 
	 files labelled 50681546 (left) and 50681549 (right). Also shown is the 
	 probability density (un-normalized) of the position of the 500 strongest 
	 spikes, for each data file. {\bf Bottom panels}: Expanded view of the 
	 probability density in the region of the central component of the integrated 
	 profiles. The abscissa are in units of the period (5757.4365 $\mu$s), while 
	 the ordinates are in arbitrary units. The horizontal bar at the top right
	 corner of the bottom two panels represents a duration of 20 $\mu$s.
	 }
\label{fig3}
\end{figure}

Then, the positions of the spikes in each file were obtained modulo the pulsar
period. A positive (negative) error in the period used will systematically 
reduce (increase) the positions of the spikes that arrive later in the data 
file. This will tend to space out the spikes in the integrated profile, which
might give the impression of an {\bf enhanced contrast} of a fringe like 
distribution of their positions. So the pulsar period was optimized so as to 
give the narrowest distribution of the positions of the spikes in each data 
file; this turned out to be 5757.436 $\pm \ 0.001 \mu$s, which is consistent 
with the value of 5757.4365 $\mu$s obtained using the prediction mode of the 
TEMPO timing package. The following results are similar irrespective of 
which period is used.

Then each of the 500 spikes was replaced by a normalized Gaussian, 
centered at the estimated position of the spike within the period, having an 
rms width equal to the estimated positional error. The resulting probability 
density of occurrence of the position of the spikes is shown in fig.~\ref{fig3} 
for the two data files mentioned above. The lower panels of fig.~\ref{fig3} 
show an expanded view of the probability density functions in the central peak 
of the integrated profile. Clearly there is no evidence for modulation by 
fringes of any periodicity; a power spectrum analysis of these profiles confirms 
the conclusion.

\section{Comparison with the method of AMDV}

The probability density function of the positions of the spikes (fig~\ref{fig3})
is the sum of several narrow Gaussian -- their widths are much smaller than
the range of abscissa that they are spread over. For a limited number of spikes
this might naturally give rise to what might appear to be a fringing over the 
broad distribution. The real issue is whether there is a persistent periodicity
in these fringes, which does not vanish as the number of spikes is increased.

\subsection{Error estimation}

The main difference between AMDV and this work is the estimation of the rms 
errors on the positions of spikes. AMDV use a uniform value of 1.7 $\mu$s 
for all 500 spikes in a file, obtained by means of ``Monte Carlo analysis''. 
This work estimates them self-consistently from the data. They range from 
about 3.2 $\mu$s to 12.3 $\mu$s in the two data files combined, with an 
average value of about 7.5 $\mu$s in each data file. Therefore AMDV's errors 
are on the average 7.5 / 1.7 $\approx$ 4.4 times smaller, and are independent
of the snr of the spikes.

Their justification is that ``Nyquist reconstruction (Bracewell 1965) allows
exact recovery of the underlying band-limited continuous function (pulsar 
signal plus noise) ...''; and therefore ``in these distributions there is 
almost no noise in the traditional sense''.

While the former statement is correct, the latter is not. Nyquist reconstruction
will not separate the signal from the noise; it will reconstruct only their
sum. So the estimation of any parameter of the signal (say, the position of a 
spike) will involve uncertainty due to the noise; the estimation error on 
such a parameter must depend upon the actual snr of the signal buried in the 
noise. Stronger spikes should have their positions estimated more accurately; 
Appendix A shows a simple method of estimating this accuracy. 

Another way of looking at this problem is the following. To obtain arbitrarily
small errors on the positions of the spikes, one needs arbitrarily large
number of ordinates, of a given snr. Now, it is true that by Nyquist 
reconstruction one can synthesize an infinite (in principle) number of ordinate 
values in any given abscissa range; and each one of them will have the same 
rms error as of the original data (Bracewell \cite{RB}). However, not all of 
the synthesized ordinates will have {\bf independent} noise; only those 
separated by the sampling interval will. So the errors on the positions of the 
spikes can not be very different from that derived in Appendix A.

This author can think of no way he could have overestimated the {\bf average} 
positional errors of the spikes by a factor of 4.4.

On the other hand, is there a possibility of {\bf underestimating} the 
positional errors of spikes? The model assumed in Appendix A is that of additive 
(system) noise. If one assumes that the pulsar signal is also noise like, for
example an amplitude modulated noise (AMN) according to Rickett \cite{BJR75}, 
one would have to add this so called {\bf self noise} (Goodman \cite{JG85}, 
Vivekanand \& Kulkarni \cite{MVSRK91}) to the system noise. However, the 
$\sigma$ of section 2 includes all forms of noise present in the data, since 
it has been estimated from the data itself, so the AMN model is unlikely to 
change the results of this work. Even then, the AMN model of the pulsar
signal may make it harder for AMDV to justify their positional errors.

\subsection{Averaging two probability distributions}

Another difference is that AMDV do not average coherently the two (presumed 
fringing) probability distributions of their fig.~2 (similar to fig.~\ref{fig3} 
here). If indeed there is a coherent periodicity in the two data files, their 
average power spectrum must enhance (or at least not reduce) the contrast of
this spectral feature, unless the periodicity is changing from file to file.

\begin{figure}
\epsfxsize=8.5cm \epsfbox{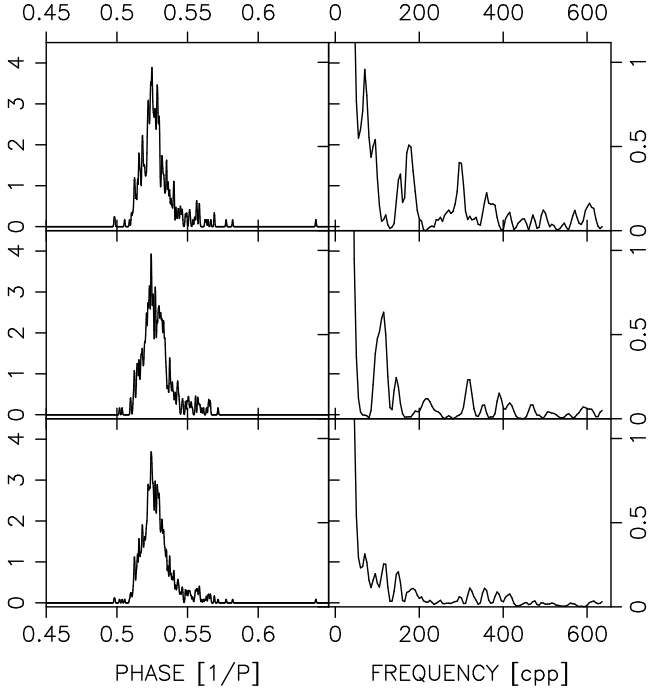}
\caption{{\bf Left Panels}: Probability density (un-normalized) of the position 
	 of the 500 strongest spikes in a data file, using a fixed error of 1.7 
	 $\mu$s for all positions; this is equal to the FWHM of 4.0 $\mu$s
	 used by AMDV. Only the distribution in the central component of the 
	 integrated profile is considered. The abscissa are in units of the 
	 period (5757.4365 $\mu$s), while the ordinates are in arbitrary units. 
	 {\bf Right Panels}: 
	 The corresponding power spectrum; the abscissa are in units of cycles 
	 per period (cpp), while the ordinates are in arbitrary units. {\bf Top 
	 pair} of figures refer to data file 50681546 while the {\bf middle 
	 pair} refer to data file 50681549; the probability density of the 
	 latter has been shifted so as to be aligned to that of the former. 
	 The {\bf bottom pair} refer to the average of the two data files.
	 }
\label{fig4}
\end{figure}

Figure~\ref{fig4} shows the probability density functions of the position of 
the spikes in the two data files, along with the corresponding power spectrum
(the probability distribution of the second file has been shifted by the 
appropriate amount to align with the first, using a cross-correlation). 
In this figure all positional errors have been set to a constant 1.7 $\mu$s, 
to recreate (if possible) the results of AMDV. There appears to be no real 
harmonic relation between the spectral features in each of the top two panels 
of fig.~\ref{fig4}. Moreover the spectra of the two files are not similar. 
This is disturbing, since the duration of each file is $\approx$ 1.16 minutes, 
and they are separated in observing time by only 3.35 minutes; a spectral 
feature that can change significantly in the latter time is unlikely to be 
considered stable in the former time, except by a special design. The bottom 
panel of fig.~\ref{fig4} shows the average of the top two probability 
distributions, and the corresponding power spectrum. The relative power in any 
spectral feature has only reduced. This remains true even for the average of 
the two power spectra (top right and middle right panels).

If at all there is a fundamental periodicity in the data, it is more 
likely to be at $\approx$ 175 (cpp), which is equal to a periodicity of 
$\approx$ 33 $\mu$s, and not at $\approx$ 20 $\mu$s obtained by AMDV. From
the width of this feature as well as the widths of its {\bf presumed}
harmonics, one is unlikely to see ``nine fairly evenly spaced sub peaks''
in the probability function as claimed by AMDV; if at all, three or four 
sub peaks are expected be seen. However, this is not the case actually.
Some low frequencies were filtered from the average (complex) spectrum and 
an inverse Fourier transform was done to bring out the quasi periodicities 
in the data, if any; nothing similar to that seen by AMDV, or similar to 
that discussed above, was observed.

Similar results were obtained when the period used in fig.~\ref{fig4} was 
the optimized value of 5757.436 $\mu$s.

Figure~\ref{fig4} was also obtained with two modifications -- using 
positional errors that were (a) half and (b) quarter of the original values 
for each spike, respectively. The corresponding probability density of 
occurrence of the position of the spikes becomes less smooth as compared 
to the bottom panels of fig.~\ref{fig3}, and similar to the left top two
panel of fig.~\ref{fig4}, but does not show any fringing. 
This is equivalent to increasing the effective signal to noise ratio of 
the data by factors 2 and 4, respectively. In other words, even if this 
work has made an error of a factor of 4 in {\bf labelling} the snr values 
of the individual samples (the relative snr values do not change), the 
results are essentially the same.

\subsection{Choice of period}

Yet another possibility is the choice of a wrong effective period by AMDV. 
They state that they used a period given by ``a computed ephemeris with 
recent timing parameters''; but they neither quote the value used nor 
mention that they optimized it as discussed above. The header in the two
data files contains the value 5757.4409 $\mu$s; this is computed by the
online software for online folding of pulsar data, and is expected to be
very accurate. In case AMDV have used this value of the period, they would 
have made an error of 4.4 nano seconds. This would lead to a systematic 
shift of approximately 12\,000 periods $\times$ 0.0044 $\mu$s $\approx$ 
53 $\mu$s between the first and last spike in the data file. Now spreading 
out of very narrow peaks along the abscissa can lead to voids in their 
distribution, and can simulate ``fringes'' in their position distribution.

\section{Acknowledgement}

I thank the referee for suggesting the AMN model, and an important
clarification regarding the period used in this paper.

\appendix

\section{Positions and errors of spikes}

Let the abscissa and the ordinate in the bottom panel of fig.~\ref{fig1} be
labelled $x$ and $y$, respectively. After trial and error, it was found that
N = 5 time samples adequately represent a typical spike, for the purpose of
estimating its position and the error on the position. The mean position was
obtained by the weighted average

\begin{eqnarray}
\phi &= \frac{\sum_{i=1}^N x_i \times y_i}{\sum_{j=1}^N y_j}\mathrm{;}
\Rightarrow  \langle \phi \rangle &\approx \frac{\sum_{i=1}^N x_i 
\times y_{0i}}{\sum_{j=1}^N y_{0j}},
\end{eqnarray}

\noindent
the third sample being at the peak of the spike; $y_{0i}$ is the true pulsar 
flux (in units of snr) while $d y_{i} = y_i - y_{0i}$ is the small variation 
due to random noise of variance 1.0 (by definition). Then the deviation $d 
\phi$ is given by

\begin{eqnarray}
\nonumber &d \phi &= \frac{\sum_{i=1}^N x_i \times d y_{i}}{\sum_{j=1}^N y_{0j}} - 
\frac{\left [ \sum_{i=1}^N x_i \times y_{0i} \right ] \times \left [ \sum_{j=1}^N 
d y_{j} \right ]}{\left [ \sum_{k=1}^N y_{0k} \right ]^2} \\
          &       &\approx \frac{\sum_{i=1}^N \left ( x_i - \langle \phi \rangle 
\right ) \times d y_{i}}{\sum_{j=1}^N y_{0j}}.
\end{eqnarray}

\noindent
Therefore the variance in the position $\sigma_\phi^2$ is given by

\begin{eqnarray}
\sigma_\phi^2 &= \langle d \phi^2 \rangle = \frac{N \times \sigma_x^2}{\left 
[ \sum_{i=1}^N y_{0i} \right ]^2}; \Rightarrow \sigma_\phi = \frac{\sqrt{N} \times 
\sigma_x}{\sum_{i=1}^N y_{0i}},
\end{eqnarray}

\noindent
where $\sigma_x$ is the standard deviation of the abscissa range used.

For a fixed $N$, and therefore a fixed denominator in eq. A3, the error
on the position $\sigma_\phi$ varies linearly with the range of abscissa
used ($\sigma_x$), as is expected. For a fixed range of abscissa, the
denominator varies as $N$, so that $\sigma_\phi$ varies inversely as 
$\sqrt{N}$; this is also expected, provided each of the $N$ values of 
the ordinate are independent measurements. The $\sigma_\phi$
is inversely proportional to the average signal to noise ratio
of the data.

To obtain an idea of the positional errors to be expected, let $N$ = 5.
For a uniform range of 5 time samples of duration 102.4 $\mu$s each, 
$\sigma_x = 5.0 \times 102.4 / \sqrt{12} \approx 147.8$ $\mu$s. The sum
of the five ordinates around the spike is 92.8 in the bottom panel of 
fig~\ref{fig1}. Therefore $\sigma_\phi = \sqrt{5} \times 147.8 / 92.8
\approx 3.6$ $\mu$s. This is the best positional error in the current
data, since this is the highest peak analyzed.

\end{document}